\documentclass[12pt]{iopart}

\usepackage{iopams}  

\usepackage{graphicx}

\newcommand{\be}{\begin{equation}}
\newcommand{\ee}{\end{equation}}

\newcommand{\trr}{{\rm Tr}}
\newcommand{\bea}{\begin{eqnarray}}
\newcommand{\eea}{\end{eqnarray}}
\newcommand{\bml}{\begin{multline}}
\newcommand{\eml}{\end{multline}}

\newcommand{\la}{\langle}

\newcommand{\ra}{\rangle}
\newcommand{\nn}{\nonumber}

\newcommand{\rectangleAA}{\begin{array}{l}\begin{picture}(40,25)

        \put (0,0) {\line(1,0){20}}
        \put (0,0) {\vector(1,0){13}}
        \put (20,0) {\line(1,0){20}}
        \put (20,0) {\vector(1,0){13}}  
        \put (40,0) {\vector(0,1){13}}
        \put (40,0) {\line(0,1){20}}
        \put (40,20) {\vector(-1,0){13}}
        \put (40,20) {\line(-1,0){20}}
        \put (20,20) {\vector(-1,0){13}}
        \put (20,20) {\line(-1,0){20}}
        \put (0,20) {\vector(0,-1){13}}
        \put (0,20) {\line(0,-1){20}}
        
        \put (4,4) {\line(1,0){16}}
        \put (4,4) {\vector(1,0){9}}
        \put (20,4) {\line(1,0){16}}
        \put (24,4) {\vector(1,0){9}}  
        \put (36,4) {\vector(0,1){9}}
        \put (36,4) {\line(0,1){12}}
        \put (36,16) {\vector(-1,0){9}}
        \put (36,16) {\line(-1,0){16}}
        \put (16,16) {\vector(-1,0){9}}
        \put (20,16) {\line(-1,0){16}}
        \put (4,16) {\vector(0,-1){9}}
        \put (4,16) {\line(0,-1){12}}

\end{picture}\end{array}}

\newcommand{\rectangleAB}{\begin{array}{l}\begin{picture}(40,25)

        \put (0,0) {\line(4,1){16}}
        \put (4,4) {\line(4,-1){16}}        
       
        \put (20,0) {\line(1,0){20}}
        \put (20,0) {\vector(1,0){13}}  
        \put (40,0) {\vector(0,1){13}}
        \put (40,0) {\line(0,1){20}}
        \put (40,20) {\vector(-1,0){13}}
        \put (40,20) {\line(-1,0){20}}
        \put (20,20) {\vector(-1,0){13}}
        \put (20,20) {\line(-1,0){20}}
        \put (0,20) {\vector(0,-1){13}}
        \put (0,20) {\line(0,-1){20}}
        
        \put (16,4) {\line(1,0){20}}
        \put (24,4) {\vector(1,0){9}}  
        \put (36,4) {\vector(0,1){9}}
        \put (36,4) {\line(0,1){12}}
        \put (36,16) {\vector(-1,0){9}}
        \put (36,16) {\line(-1,0){16}}
        \put (16,16) {\vector(-1,0){9}}
        \put (20,16) {\line(-1,0){16}}
        \put (4,16) {\vector(0,-1){9}}
        \put (4,16) {\line(0,-1){12}}

\end{picture}\end{array}}

\newcommand{\rectangleAC}{\begin{array}{l}\begin{picture}(40,25)
      \put (0,0) {\line(1,0){20}}
      \put (0,0) {\vector(1,0){13}}
      \put (20,0) {\line(1,0){20}}
      \put (20,0) {\vector(1,0){13}}  
      \put (40,0) {\vector(0,1){13}}
      \put (40,0) {\line(0,1){20}}
      \put (40,20) {\vector(-1,0){13}}
      \put (40,20) {\line(-1,0){20}}
      \put (20,20) {\vector(-1,0){13}}
      \put (20,20) {\line(-1,0){20}}
      \put (0,20) {\vector(0,-1){13}}
      \put (0,20) {\line(0,-1){20}}   
      \put (21.5,4) {\line(-1,4){3}}
      \put (18.5,4) {\line(1,4){3}} 
      \put (4,4) {\line(1,0){14.5}}
      \put (4,4) {\vector(1,0){9}}
      \put (36,4) {\line(-1,0){14.5}}
      \put (36,4) {\vector(-1,0){9}}  
      \put (36,16) {\vector(0,-1){9}}
      \put (36,16) {\line(0,-1){12}}
      \put (21.5,16) {\vector(1,0){11.5}}
      \put (21.5,16) {\line(1,0){14.5}}
      \put (18.5,16) {\vector(-1,0){11.5}}
      \put (18.5,16) {\line(-1,0){14.5}}
      \put (4,16) {\vector(0,-1){9}}
      \put (4,16) {\line(0,-1){12}}

\end{picture}\end{array}}

\newcommand{\rectangleAD}{\begin{array}{l}\begin{picture}(40,25)
      \put (0,0) {\line(1,0){15}}
      \put (0,0) {\vector(1,0){13}}
      \put (40,0) {\line(-1,0){15}}
      \put (40,0) {\vector(-1,0){13}}  
      \put (40,20) {\vector(0,-1){13}}
      \put (40,20) {\line(0,-1){20}}
      \put (25,20) {\vector(1,0){8}}
      \put (25,20) {\line(1,0){15}}
      \put (15,20) {\vector(-1,0){8}}
      \put (15,20) {\line(-1,0){15}}
      \put (0,20) {\vector(0,-1){13}}
      \put (0,20) {\line(0,-1){20}}   
      \put (21.5,4) {\line(-1,4){3}}
      \put (18.5,4) {\line(1,4){3}} 
      \put (25,0) {\line(-1,2){10}}
      \put (15,0) {\line(1,2){10}}
 
      \put (4,4) {\line(1,0){14.5}}
      \put (4,4) {\vector(1,0){9}}
      \put (36,4) {\line(-1,0){14.5}}
      \put (36,4) {\vector(-1,0){9}}  
      \put (36,16) {\vector(0,-1){9}}
      \put (36,16) {\line(0,-1){12}}
      \put (21.5,16) {\vector(1,0){11.5}}
      \put (21.5,16) {\line(1,0){14.5}}
      \put (18.5,16) {\vector(-1,0){11.5}}
      \put (18.5,16) {\line(-1,0){14.5}}
      \put (4,16) {\vector(0,-1){9}}
      \put (4,16) {\line(0,-1){12}}

\end{picture}\end{array}}
\newcommand{\rectangleAE}{\begin{array}{l}\begin{picture}(40,25)

        \put (0,0) {\line(1,0){20}}
        \put (0,0) {\vector(1,0){13}}
        \put (20,0) {\line(1,0){20}}
        \put (20,0) {\vector(1,0){13}}  
        \put (40,0) {\vector(0,1){13}}
        \put (40,0) {\line(0,1){20}}
        \put (40,20) {\vector(-1,0){13}}
        \put (40,20) {\line(-1,0){16}}
        \put (24,20) {\vector(0,-1){13}}
        \put (24,20) {\line(0,-1){16}}
        \put (24,4) {\line(1,0){12}}
        \put (24,4) {\vector(1,0){9}} 

        \put (16,20) {\vector(-1,0){9}}
        \put (16,20) {\line(-1,0){16}}
        \put (0,20) {\vector(0,-1){13}}
        \put (0,20) {\line(0,-1){20}}

        \put (4,4) {\line(1,0){12}}
        \put (4,4) {\vector(1,0){9}}
        \put (16,4) {\line(0,1){16}}
        \put (16,4) {\vector(0,1){9}}

        \put (36,4) {\vector(0,1){9}}
        \put (36,4) {\line(0,1){12}}
        \put (36,16) {\vector(-1,0){9}}
        \put (36,16) {\line(-1,0){16}}
        \put (16,16) {\vector(-1,0){9}}
        \put (20,16) {\line(-1,0){16}}
        \put (4,16) {\vector(0,-1){9}}
        \put (4,16) {\line(0,-1){12}}

\end{picture}\end{array}}

\newcommand{\rectangleAF}{\begin{array}{l}\begin{picture}(40,25)

        \put (0,0) {\line(1,0){17}}
        \put (0,0) {\vector(1,0){13}}
        \put (40,0) {\line(-1,0){16}}
        \put (40,0) {\vector(-1,0){13}}  
        \put (40,20) {\vector(0,-1){13}}
        \put (40,20) {\line(0,-1){20}}
        \put (24,20) {\vector(1,0){9}}
        \put (24,20) {\line(1,0){16}}
       
        \put (36,4) {\line(-1,0){14}}
        \put (36,4) {\vector(-1,0){9}} 

        \put (16,20) {\vector(-1,0){9}}
        \put (16,20) {\line(-1,0){16}}
        \put (0,20) {\vector(0,-1){13}}
        \put (0,20) {\line(0,-1){20}}

        \put (4,4) {\line(1,0){12}}
        \put (4,4) {\vector(1,0){9}}
        \put (16,4) {\line(0,1){16}}
        \put (16,4) {\vector(0,1){9}}

        \put (36,16) {\vector(0,-1){9}}
        \put (36,16) {\line(0,-1){12}}
        \put (24,16) {\vector(1,0){9}}
        \put (24,16) {\line(1,0){12}}
        \put (16,16) {\vector(-1,0){9}}
        \put (18,16) {\line(-1,0){14}}
        \put (4,16) {\vector(0,-1){9}}
        \put (4,16) {\line(0,-1){12}}
        \put (24,0) {\vector(0,1){13}}
        \put (24,0) {\line(0,1){16}}
        
        \put (17.25,0) {\line(1,3){6.65}}
        \put (22,4) {\line(-1,3){4}}
\end{picture}\end{array}}

\newcommand{\rectangleAG}{\begin{array}{l}\begin{picture}(40,25)
      \put (4,0) {\line(1,0){16}}
      \put (4,0) {\vector(1,0){9}}
      \put (20,0) {\line(1,0){20}}
      \put (20,0) {\vector(1,0){13}}  
      \put (40,0) {\vector(0,1){13}}
      \put (40,0) {\line(0,1){20}}
      \put (40,20) {\vector(-1,0){13}}
      \put (40,20) {\line(-1,0){20}}
      \put (20,20) {\vector(-1,0){13}}
      \put (20,20) {\line(-1,0){20}}
      \put (0,20) {\vector(0,-1){13}}
      \put (0,20) {\line(0,-1){16}}   
      \put (21.5,4) {\line(-1,4){3}}
      \put (18.5,4) {\line(1,4){3}} 
      \put (0,4) {\line(1,0){18.5}}
      \put (0,4) {\vector(1,0){13}}
      \put (36,4) {\line(-1,0){14.5}}
      \put (36,4) {\vector(-1,0){9}}  
      \put (36,16) {\vector(0,-1){9}}
      \put (36,16) {\line(0,-1){12}}
      \put (21.5,16) {\vector(1,0){11.5}}
      \put (21.5,16) {\line(1,0){14.5}}
      \put (18.5,16) {\vector(-1,0){11.5}}
      \put (18.5,16) {\line(-1,0){14.5}}
      \put (4,16) {\vector(0,-1){9}}
      \put (4,16) {\line(0,-1){16}}

\end{picture}\end{array}}

\newcommand{\rectangleAH}{\begin{array}{l}\begin{picture}(40,25)
    \put (0,0) {\line(1,0){17}}
        \put (0,0) {\vector(1,0){13}}
        \put (40,0) {\line(-1,0){16}}
        \put (40,0) {\vector(-1,0){13}}  
        \put (40,20) {\vector(0,-1){13}}
        \put (40,20) {\line(0,-1){20}}
        \put (24,20) {\vector(1,0){9}}
        \put (24,20) {\line(1,0){16}}
       
        \put (22,4) {\line(1,0){14}}
        \put (22,4) {\vector(1,0){11}} 

        \put (16,20) {\vector(-1,0){9}}
        \put (16,20) {\line(-1,0){16}}
        \put (0,20) {\vector(0,-1){13}}
        \put (0,20) {\line(0,-1){20}}
         
        \put (36,4) {\vector(0,1){9}}
        \put (36,4) {\line(0,1){12}}
        \put (4,16) {\vector(1,0){9}}
        \put (4,16) {\line(1,0){14}}
        \put (4,4) {\vector(0,1){9}}
        \put (4,4) {\line(0,1){12}}
       
        \put (17.25,0) {\line(1,3){6.65}}
        \put (22,4) {\line(-1,3){4}}
        \put (24,0) {\line(-5,1){20}}
        \put (36,16) {\line(-5,1){20}}
\end{picture}\end{array}}
\newcommand{\rectangleAI}{\begin{array}{l}\begin{picture}(40,25)
    \put (0,0) {\line(1,0){18}}
        \put (0,0) {\vector(1,0){13}}
        \put (40,0) {\line(-1,0){18}}
        \put (40,0) {\vector(-1,0){13}}  
        \put (40,16) {\vector(0,-1){9}}
        \put (40,16) {\line(0,-1){16}}
        \put (22,20) {\vector(1,0){11}}
        \put (22,20) {\line(1,0){14}}
       
        \put (36,4) {\line(-1,0){13}}
        \put (36,4) {\vector(-1,0){9}} 
        \put (23,16) {\line(1,0){17}}
        \put (23,16) {\vector(1,0){10}}
        
        \put (17.5,20) {\vector(-1,0){8.5}}
        \put (17.5,20) {\line(-1,0){17.5}}
        \put (0,20) {\vector(0,-1){13}}
        \put (0,20) {\line(0,-1){20}}
         
        \put (36,20) {\vector(0,-1){13}}
        \put (36,20) {\line(0,-1){16}}
        \put (16.5,16) {\vector(-1,0){9.5}}
        \put (16.5,16) {\line(-1,0){12.5}}
        \put (4,16) {\vector(0,-1){9}}
        \put (4,16) {\line(0,-1){12}}
        \put (4,4) {\vector(1,0){9}}
        \put (4,4) {\line(1,0){13}}

        \put (18,0) {\line(1,3){5.4}}
        \put (23,4) {\line(-1,3){5.4}}
        \put (22,0) {\line(-1,3){5.4}}
       \put (17,4) {\line(1,3){5.4}}
\end{picture}\end{array}}

\newcommand{\rectangleAAd}{\begin{array}{l}\begin{picture}(40,25)

        \put (0,0) {\line(1,0){20}}
        \put (0,0) {\vector(1,0){13}}
        \put (20,0) {\line(1,0){20}}
        \put (20,0) {\vector(1,0){13}}  
        \put (40,0) {\vector(0,1){13}}
        \put (40,0) {\line(0,1){20}}
        \put (40,20) {\vector(-1,0){13}}
        \put (40,20) {\line(-1,0){20}}
        \put (20,20) {\vector(-1,0){13}}
        \put (20,20) {\line(-1,0){20}}
        \put (0,20) {\vector(0,-1){13}}
        \put (0,20) {\line(0,-1){20}}
        
        \put (4,4) {\line(0,1){12}}
        \put (4,4) {\vector(0,1){8}}
        \put (20,4) {\line(-1,0){16}}
        \put (16,4) {\vector(-1,0){8}}  
        \put (36,4) {\vector(-1,0){8}}
        \put (36,4) {\line(-1,0){16}}
        \put (36,16) {\vector(0,-1){8}}
        \put (36,16) {\line(0,-1){12}}
        \put (24,16) {\vector(1,0){8}}
        \put (20,16) {\line(1,0){16}}
        \put (4,16) {\vector(1,0){8}}
        \put (4,16) {\line(1,0){16}}

\end{picture}\end{array}}

\newcommand{\eqn}[1]{Eq.~(\ref{#1})}
\newcommand{\Eqn}[1]{Eq.~(\ref{#1})}
\newcommand{\eqns}[2]{Eqs.~(\ref{#1}) and (\ref{#2})}
\newcommand{\eqnss}[2]{Eqs.~(\ref{#1}---\ref{#2})}
\newcommand{\eqnsss}[3]{Eqs.~(\ref{#1}), (\ref{#2}) and (\ref{#3})}

\newcommand{\sect}[1]{Section~\ref{#1}}

\begin{document}

\title[Integrals over SU($N$)]{Integrals over SU($N$)}

\author{Jesse Carlsson}

\ead{jesse.carlsson@gmail.com}
\begin{abstract}
In this paper we calculate a number of integrals over SU($N$) of interest in Hamiltonian Lattice Gauge Theory calculations.
\end{abstract}

\maketitle

\section{Introduction: file preparation and submission}
This paper is concerned with the analytic calculation of certain group integrals over SU($N$) which are of interest in Hamiltonian Lattice Gauge Theory (LGT)~\cite{Kogut:1975ag}. In particular, the integrals in question arise in calculations of ground state energies and mass gaps using the plaquette expansion method~\cite{Hollenberg:1993bp}. While these integrals have traditionally been handled with with numerical Monte Carlo integration, in this paper we demonstrate how these integrals may be calculated analytically. \\

Traditionally Hamiltonian LGT techniques have been restricted to simple ``trial'' states (most commonly the strong coupling ground state which is anihilated by the chromo-electric field). The techniques presented in this paper are are of use in the extension of standard Hamiltonian LGT techniques beyond trivial ``trial'' states, at least in 2+1 dimensions, to include more complicated trial states than the stong coupling ground state. For example, the techniques presented in \sect{preliminaries} have been applied in variational estimates of glueball masses in 2+1 dimensions~\cite{Carlsson:2002ss,Carlsson:2003wx}. It is thought that similar techniques may be of use in plaquette expansion method calculations of glueball masses in 2+1 dimensions. \\

Much work has been carried out on the topic of integration over the classical
compact groups. The
subject has been studied in great depth in the context of random
matrices and combinatorics. Many analytic results in terms of determinants are
available for integrals of various functions over unitary, orthogonal 
and symplectic groups~\cite{Baik:2001}. Unfortunately similar results
for SU($N$) are not to our knowledge available. 
Their primary use has been in the study of Ulam's problem concerning
the distribution of the length of 
the longest increasing subsequence in permutation
groups~\cite{Rains:1998,Widom:2001}. Connections between random
permutations and Young tableaux~\cite{Regev:1981} allow an interesting
approach to combinatorial problems involving Young tableaux. A 
problem of particular interest is the counting of Young tableaux
of bounded height~\cite{Gessel:1990} which is closely related to the
problem of counting singlets in product representations. Group integrals similar to those examined in
this paper have also appeared in studies of the distributions of the
eigenvalues of random matrices~\cite{Diaconis:1994,Widom:1999}. \\

In the last 20 years it appears that not much effort has been devoted to the
subject of group integration in the context of LGT. The last
significant development was due to Creutz who
developed a diagrammatic technique for calculating specific SU($N$)
integrals~\cite{Creutz:1978ub} using link variables. 
This technique allows strong coupling
matrix elements to be calculated for SU($N$)~\cite{ConradPhD}
and has more recently been used in the loop formulation of quantum
gravity where spin networks are of interest~\cite{DePietri:1997pj,Rovelli:1995ac,Ezawa:1997bv}.\\

This paper is structured as follows. \sect{preliminaries} contains calculations of some general SU($N$) integrals which are used in later sections. \sect{creutzextension} contains integrals of SU($N$) group elements which are then applied in \sect{momentints} to the calculation of a number of integrals that arise in the plaquette expansion method in 2+1 dimensional Hamiltonian LGT using a one plaquette trial state. 

\section{Preliminaries}
\label{preliminaries}
In an earlier paper~\cite{Carlsson:2003wx} a useful technique for performing SU($N$)
integrals was described. In this paper the following results in terms of Toeplitz determinants were derived,
\bea
\fl G_{{\rm U}(N)}(c,d) &= \int_{{\rm U}(N)} dU e^{c {\rm Tr} U + d {\rm Tr} U^\dagger} = \det\left[ I_{j-i}\left(2\sqrt{c d}\right)\right]_{1\le i,j\le N}\label{simplegen} \\
\fl G_{{\rm SU}(N)}(c,d) &=\int_{{\rm SU}(N)} dU e^{c {\rm Tr} U + d {\rm Tr} U^\dagger} =
\sum_{l=-\infty}^{\infty} \left(\frac{d}{c}\right)^{l N/2}
\det \left[ I_{l+j-i}\left(2\sqrt{cd}\right)\right]_{i\le i,j \le N}.
\label{coolsum}
\eea
Here $c$ is any number, $dU$ is the Haar measure normalised so that $\int dU = 1$, the quantities inside the determinant are to be interpreted as the $(i,j)$-th entry of an $N\times N$ matrix, and $I_n$ is the $n$-th order modified Bessel function of the first kind defined, for integers $n$, by
\bea
I_n(2 x) = \sum_{k=0}^{\infty} \frac{x^{2k+n}}{k!(k+n)!}.
\eea

\Eqn{coolsum} can be used to as a generating function to calculate integrals of the form
\bea
 \int_{{\rm SU}(N)} (\trr U)^n (\trr U^\dagger)^m   dU e^{c {\rm Tr} U + d {\rm Tr} U^\dagger},
\eea
by direct differentiation. This process allows easy calculation of integrals of polynomials in $\trr U$ and $\trr U^\dagger$. It does not allow the simple calculation of integrals involving the more complicated trace variables, $(\trr U^n)^m$ for integers $n$ and $m$. 

Making use of the same technique described in~\cite{Carlsson:2003wx}, the following more generals integrals can be calculated, 
\bea
 \int_{{\rm U}(N)}dU \chi_{n_1 n_2 \cdots n_N}(U)e^{c {\rm Tr} U + d {\rm Tr} U^\dagger} = \left(\frac{d}{c}\right)^{\sum_l n_l/2}\det I_{j-i+n_i}(2\sqrt{cd})
\label{prelimdefU} \hspace{0.2cm} {\rm and}\\
\fl \int_{{\rm SU}(N)}dU \chi_{n_1 n_2 \cdots n_N}(U)e^{c {\rm Tr} U + d {\rm Tr} U^\dagger}= \left(\frac{d}{c}\right)^{\sum_k n_k/2}\sum_{l=-\infty}^{\infty}\left(\frac{d}{c}\right)^{Nl/2}\det I_{j-i+l+n_i}(2\sqrt{cd}).
\label{prelimdefSU}
\eea
To arrive at these results, instead of~Eq.~(15) in~\cite{Carlsson:2003wx}, we make use of the following result due to Weyl~\cite{Weyl:1946}(with implicit sums over repeated indices understood) and follow the same procedure,
\bea
\fl \Delta(\phi_1,\ldots,\phi_N)\chi_{n_1\cdots n_N} &=& \frac{1}{\sqrt{N!}} \varepsilon_{i_1 i_2\cdots i_N} e^{i
\phi_1(N-i_1+n_{i_1})}e^{i
\phi_2(N-i_2+n_{i_2})} \cdots  e^{i
\phi_N(N-i_N+n_{i_N})}.
\label{character-Weyl}
\eea
Here $\varepsilon_{i_1\ldots i_n}$ is the totally antisymmetric
Levi-Civita tensor defined to be 1 if $\{i_1,\ldots,i_n\}$ is an even
permutation of $\{1,2,\ldots,n\}$, $-1$ if it is an odd permutation
and 0 otherwise (i.e. if an index is repeated).

\eqns{prelimdefU}{prelimdefSU} allow the simple calculation of many integrals involving $(\trr U^n)^m$ for integers $n$ and $m$. To perform these calculations expressions of the form $(\trr U^n)^m$ must be written in terms of characters. To do this the following results are useful \cite{Weyl:1946,Bars:1980yy},
\bea
\chi_{r_1r_2\ldots r_{N-1}}(U) = \det\left[ \chi_{r_i+i-j}\right]_{1\le i,j\le N-1}\hspace{0.3cm} {\rm , where}\\
\chi_n(U)= \sum_{k_1, \ldots ,k_n} \delta\left(\sum_{i=1}^n i k_i - n\right)\prod_{j=1}^n\frac{1}{k_j!j^{k_j}}(\trr U^j)^{k_j}.
\eea  
Using these results, expressions for characters in terms of trace variables can be calculated. These results can be rearranged to give expressions for trace variables in terms of characters. As an example, for SU($N$) it can be shown that 
\bea
(\trr U)^2 &=& \chi_2(U) + \chi_{11}(U) \hspace{0.2cm} {\rm and} \label{truchar1}\\
\trr (U^2) &=& \chi_2(U) - \chi_{11}(U). \label{truchar2}
\eea
Making use of \eqnsss{truchar1}{truchar2}{prelimdefSU} the following integrals can be easily calculated
\bea
\int_{{\rm SU}(N)}dU \trr (U^2) e^{c {\rm Tr} U + d {\rm Tr} U^\dagger} \hspace{0.2cm}{\rm and} \hspace{0.2cm} \int_{{\rm SU}(N)}dU (\trr U)^2 e^{c {\rm Tr} U + d {\rm Tr} U^\dagger}.
\eea 
\section{Creutz's Method}
\label{creutzextension}

In this section we make use of Creutz's method for calculating strong coupling integrals over SU($N$) to calculate more complicated integrals. In the terminology of Hamiltonian LGT, we use Creutz's method to calculate one-plaquette trial state integrals. The integrals we consider in this section are of use in the calculation of matrix elements arising in Hamiltonian LGT, particularly in calculations involving Hamiltonian moments.

The first two integrals we consider calculate in this section are,
\bea
I_{ab;a'b'}(c) &=& \int dU e^{c\trr(U+U^\dagger)}U_{ab}U_{a'b'}\hspace{0.5cm} {\rm and}\\
J_{ab;a'b'}(c) &=&  \int dU e^{c\trr(U+U^\dagger)}U_{ab}(U^\dagger)_{a'b'},\label{J}
\eea
where $c$ is a complex valued variable.

Let us start with $I_{ab;a'b'}(c)$. Expanding the exponential in a Taylor series and also expanding the resulting binomial terms gives,
\bea
\fl I_{ab;a'b'}(c)&=& \sum_{m=0}^\infty\sum_{i=0}^{m} \left(\begin{array}{c}m\\i\end{array}\right)\frac{c^m}{m!}\int dU U_{a_1a_1}\cdots U_{a_{m-i}a_{m-i}}(U^\dagger)_{b_1b_1}\cdots (U^\dagger)_{b_ib_i}U_{ab}U_{a'b'}\label{Iaba'b'}.
\eea
Following the diagrammatic procedure first described by Creutz~\cite{Creutz:1978ub}, the contributing integrals within the sum reduce to products of paired Levi-Civita symbols. The indices of each pair of Levi-Civita symbols are completely contracted with one another with the exception of those that contain an uncontracted $U$ index $a$, $b$, $a'$ or $b'$. Following Creutz's procedure, each non-zero integral under the sum in \eqn{Iaba'b'} can be reduced to linear combinations of the following products of Levi-Civita symbols,
\bea
\epsilon_{a A_1 A_2 \cdots A_{N-1}}\epsilon_{b A_1 A_2 \cdots A_{N-1}}\epsilon_{a' B_1 B_2 \cdots B_{N-1}}\epsilon_{b' B_1 B_2 \cdots B_{N-1}}\propto \delta_{ab}\delta_{a'b'}, {\rm and}\label{creutzproc1}\\
\epsilon_{a a' A_1 A_2 \cdots A_{N-2}}\epsilon_{b b' A_1 A_2 \cdots A_{N-2}} \propto  \delta_{ab}\delta_{a'b'}-\delta_{a b'}\delta_{b a'},\label{creutzproc2} 
\eea
as the paired Levi-Civita symbols can contract with uncontracted $U$ index pairs, $(a,b)$ and $(a',b')$ in two possible ways, one involving a single pair of Levi-Civita symbols (\eqn{creutzproc1}) and the other involving two (\eqn{creutzproc2}). The additional, fully contracted Levi-Civita symbols contrbute a multiplicative factor to each term. The simplification to delta functions presented in \eqns{creutzproc1}{creutzproc2} is easily obtained from the standard result from differential geometry,
\be
\epsilon_{a_1 a_2 \cdots a_n} \epsilon_{b_1 b_2 \cdots b_n} = \det\left(\delta_{a_i b_j} \right)_{1\le i, j \le n}. \label{diffgeom}
\ee   
Hence, $I_{ab;a'b'}$ may be expressed as a linear combination of delta functions,
\bea
I_{ab;a'b'}(c) = b_1(c) \delta_{ab}\delta_{a'b'} + b_2(c) \delta_{ab'}\delta_{ba'},\label{lincomb}
\eea  
where $b_1$ and $b_2$ are functions of $c$ to be determined. To find these functions we contract over indices in \eqn{lincomb} to contruct equations for $b_1$ and $b_2$ in terms of known integrals. For this particular integral we construct two equations for $b_1$ and $b_2$ by firstly, mulitpling both sides of \eqn{lincomb} by $\delta_{ba}\delta_{b'a'}$ and summing over repeated indices and secondly, mulitplying both sides of \eqn{lincomb} by $\delta_{a'b}\delta_{b'a}$ and summing over repeated indices. These operations result in the following two equations:
\bea
I_{aa;bb}(c) &=& N^2 b_1(c) + N b_2(c)\\
I_{ab;ba}(c) &=& N b_1(c)+ N^2 b_2(c),
\eea 
which can be solved for $b_1$ and $b_2$ and substituted in \eqn{lincomb} to give our final result for $I_{ab;a'b'}$ in terms of known integrals,
\bea
I_{ab;a'b'}(c) &=&  \frac{\delta_{ab}\delta_{a'b'}}{N^2-1} \int dU e^{c\trr(U+U^\dagger)}\left[(\trr U)^2 - \frac{\trr(U^2)}{N}\right] \\
&&+\frac{\delta_{ab'}\delta_{ba'}}{N^2-1} \int dU e^{c\trr(U+U^\dagger)}\left[\trr(U^2) - \frac{(\trr U)^2}{N}\right].\label{b2}
\eea

Our approach to calculating $J_{ab;a'b'}$, defined in \eqn{J}, follows the approach for $I_{ab;a'b'}$ decribed above. As was done for $I_{ab;a'b'}$, it can be shown that $J_{ab;a'b'}$ may be expressed as a linear combination of delta functions using the method due to Creutz,
\bea
J_{ab;a'b'}(c) &=& d_1(c) \delta_{ab}\delta_{a'b'} + d_2(c) \delta_{ab'}\delta_{ba'}. 
\label{Jindels}
\eea
By contracting with appropriate delta functions we arrive at simultaneous equations for $d_1$ and $d_2$, which are solved to give an expression for $J_{ab;a'b'}$ in terms of known integrals
\bea
J_{ab;a'b'}(c) &=& \frac{\delta_{ab}\delta_{a'b'}}{N^2-1} \int dU e^{c\trr(U+U^\dagger)}\left(\trr U \trr U^\dagger - 1 \right)\\
&& + \frac{\delta_{ab'}\delta_{ba'}}{N^2-1} \int\!\! dU e^{c\trr(U+U^\dagger)}\left(N - \frac{1}{N}\trr U \trr U^\dagger\right). \label{Jdetail}
\eea
This result agrees with the strong coupling result $J_{ab;a'b'}(0) = \delta_{ab'}\delta_{ba'}/N$ since
\bea
 \int dU \trr U \trr U^\dagger = \int dU  = 1.
\eea

This procedure can be used to evaluate more complicated integrals. For example, following Creutz's method again,
\be
I_{ab;a'b';a''b''}(c) = \int dU e^{c\trr(U+U^\dagger)}U_{ab}U_{a'b'}U_{a''b''},
\label{i3defn}
\ee
can be written as as linear combinations of the following products of Levi-Civita symbols,
\bea
\fl \epsilon_{a A_1 \cdots A_{N-1}}\epsilon_{b A_1 \cdots A_{N-1}}\epsilon_{a' B_1\cdots B_{N-1}}\epsilon_{b' B_1 \cdots B_{N-1}}\epsilon_{a'' C_1 \cdots C_{N-1}}\epsilon_{b'' C_1 \cdots C_{N-1}}\propto  \delta_{ab}\delta_{a'b'}\delta_{a''b''},\\\label{creutzproc3}
\fl \epsilon_{a a' A_1 \cdots A_{N-2}}\epsilon_{b b' A_1 \cdots A_{N-2}}\epsilon_{a'' B_1 \cdots B_{N-1}}\epsilon_{b'' B_1 \cdots B_{N-1}}\propto (\delta_{ab}\delta_{a'b'}-\delta_{a b'}\delta_{b a'})\delta_{a'' b''},\\\label{creutzproc4} 
\fl \epsilon_{a a'' A_1 \cdots A_{N-2}}\epsilon_{b b'' A_1 \cdots A_{N-2}}\epsilon_{a' B_1 \cdots B_{N-1}}\epsilon_{b' B_1 \cdots B_{N-1}} \propto (\delta_{ab}\delta_{a''b''}-\delta_{a b''}\delta_{b a''})\delta_{a' b'},\\\label{creutzproc5}
\fl \epsilon_{a' a'' A_1 \cdots A_{N-2}}\epsilon_{b' b'' A_1 \cdots A_{N-2}}\epsilon_{a B_1 \cdots B_{N-1}}\epsilon_{b B_1 \cdots B_{N-1}} \propto
(\delta_{a'b'}\delta_{a''b''}-\delta_{a' b''}\delta_{b' a''})\delta_{a b}, \quad {\rm and} \label{creutzproc6}\\
\fl \epsilon_{a a' a'' A_1 \cdots A_{N-3}}\epsilon_{b b' b'' A_1 \cdots A_{N-3}} \propto 
\delta_{a b''} (\delta_{a'b'} \delta_{a''b} - \delta_{a'b} \delta_{a''b'}) + 
    \delta_{ab'}(-\delta_{a'b''}\delta_{a''b}+ \delta_{a'b}\delta_{a''b''}) \nn\\
 + 
    \delta_{ab} (\delta_{a'b''}\delta_{a''b'} - \delta_{a'b'}\delta_{a''b''}).\label{creutzproc7}
\eea
These results are easily derived using \eqn{diffgeom}. Hence, $I_{ab;a'b';a''b''}$ may be written as the following linear combination
\bea
I_{ab;a'b';a''b''}(c) &=& x_1 \delta_{a b''} \delta_{a'b'} \delta_{a''b} + x_2 \delta_{a b''}\delta_{a'b} \delta_{a''b'} + x_3 
    \delta_{ab'}\delta_{a'b''}\delta_{a''b} \nn\\
&&+  x_4\delta_{ab'}\delta_{a'b}\delta_{a''b''} + x_5 
    \delta_{ab} \delta_{a'b''}\delta_{a''b'} +x_6 \delta_{ab} \delta_{a'b'}\delta_{a''b''}. \label{i3lincomb}
\eea
A collection of six equations for the $x_i$ coefficients may be constructed by contracting appropriate indices as follows:
\bea
\left(\begin{array}{cccccc} 
 N^2 & N   & N   & N^2 & N^2 & N^3 \\
 N   & N^2 & N^2 & N   & N^3 & N^2 \\
 N^3 & N^2 & N^2 & N   & N   & N^2 \\
 N   & N^2 & N^2 & N^3 & N   & N^2 \\
 N^2 & N^3 & N   & N^2 & N^2 & N \\
 N^2 & N   & N^3 & N^2 & N^2 & N                  
     \end{array}\right)\left(\begin{array}{c} x_1\\x_2\\x_3\\x_4\\x_5\\x_6     
     \end{array}\right)
= \left(\begin{array}{c} 
I_{aa;bb;cc}\\I_{aa;bc;cb}\\I_{ab;cc;ba}\\I_{ab;ba;cc}\\I_{ab;bc;ca}\\
I_{ab;ca;bc}          
     \end{array}\right)\equiv
\left(\begin{array}{c} 
      i_1\\i_2\\i_3\\i_4\\i_5\\i_6       
     \end{array}\right) \label{idefs}
\eea
This system may be solved to give,
\bea
x_1&=&\frac{2i_2 + 2i_4 -N(i_1 + i_5 + i_6)+(N^2-2) i_3}{(N-2)(N-1)N(N+1)(N+2)}\label{x1}\\
x_2&=& \frac{2i_1 + 2i_6 -N(i_2 + i_3 + i_4)+(N^2-2) i_5}{(N-2)(N-1)N(N+1)(N+2)}\\
x_3&=& \frac{2i_1 + 2i_5 -N(i_2 + i_3 + i_4)+(N^2-2) i_6}{(N-2)(N-1)N(N+1)(N+2)}\\
x_4&=& \frac{2i_2 + 2i_3 -N(i_1 + i_5 + i_6)+(N^2-2) i_4}{(N-2)(N-1)N(N+1)(N+2)}\\
x_5 &=& \frac{2i_3 + 2i_4 -N(i_1 + i_5 + i_6)+(N^2-2) i_2}{(N-2)(N-1)N(N+1)(N+2)}\\
x_6 &=& \frac{2i_5 + 2i_6 -N(i_2 + i_3 + i_4)+(N^2-2) i_1}{(N-2)(N-1)N(N+1)(N+2)}\label{x6}
\eea
Making use of the definitions in \eqn{idefs}, we see that $i_2 = i_3 = i_4$ and $i_5 = i_6$. Making these replacements in \eqnss{x1}{x6} and substituting the results in \eqn{i3lincomb} gives,
\bea
\fl I_{ab;a'b';a''b''}(c) = \frac{-N(i_1 + 2 i_5)+(N^2+2) i_2}{N(N^2-1)(N^2-4)}(\delta_{a b''} \delta_{a'b'} \delta_{a''b}+\delta_{ab'}\delta_{a'b}\delta_{a''b''} + \delta_{ab} \delta_{a'b''}\delta_{a''b'}) \nn\\
+  \frac{2i_1 -3 N i_2 + N^2 i_5}{N(N^2-1)(N^2-4)} (\delta_{a b''}\delta_{a'b} \delta_{a''b'} + \delta_{ab'}\delta_{a'b''}\delta_{a''b})\nn\\
+ \frac{4i_5 -3 N i_2 +(N^2-2) i_1}{N(N^2-1)(N^2-4)}\delta_{ab} \delta_{a'b'}\delta_{a''b''}.\label{I3}
\eea
The integrals, $i_1$, $i_2$ and $i_5$ can all be calculated using the procedure set out in \sect{preliminaries}.\\

In order to calculate all integrals of three $U_{ij}$ or $(U^\dagger)_{kl}$ factors only one more integral is required,
\bea
J_{ab;a'b';a''b''}(c) &=& \int dU e^{c\trr(U+U^\dagger)}U_{ab}(U^\dagger)_{a'b'}U_{a''b''}.
\eea
Other integrals of three group elements can be calculated using results for $I_{ab;a'b';a''b''}$ and $J_{ab;a'b';a''b''}$ by making appropriate changes of variables. 

$J_{ab;a'b';a''b''}$ can be calculated in the same way as $I_{ab;a'b';a''b''}$, in fact $J_{ab;a'b';a''b''}$ is a linear combination of the same delta functions as $I_{ab;a'b';a''b''}$ with coefficients defined by \eqnss{x1}{x6} but with the definitions of $i_1, \ldots, i_6$ modified to reflect the integral in question (i.e. $i_1 = J_{aa;bb;cc}, \ldots ,i_6 = J_{ab,ca;ba}$). The final result can be shown to be
\bea
\fl J_{ab;a'b';a''b''}(c) = \frac{2Ni_5 -N i_1 + (N^2-2) i_3}{N(N^2-1)(N^2-4)} \delta_{a b''} \delta_{a'b'} \delta_{a''b} \nn\\
+\frac{2i_1 -N i_3 - N^2 i_5}{N(N^2-1)(N^2-4)}(
 \delta_{a b''}\delta_{a'b} \delta_{a''b'} + 
    \delta_{ab'}\delta_{a'b''}\delta_{a''b}) \nn\\
+ \frac{ 2i_3 - N i_1 + N(N^2-2)i_5}{N(N^2-1)(N^2-4)}( \delta_{ab'}\delta_{a'b}\delta_{a''b''} +   \delta_{ab} \delta_{a'b''}\delta_{a''b'}) \nn\\
+\frac{2(2-N^2)i_5 -N i_3 +(N^2-2) i_1}{N(N^2-1)(N^2-4)} \delta_{ab} \delta_{a'b'}\delta_{a''b''} . \label{JJ3}
\eea
Using a software package cabable of symbolic manipulation, such as Mathematica, this technique can be easily extended to more complicated integrals, involving four or more $U$ and / or $U^\dagger$ group elements in the integrand. In order to accelerate the calculation of such integrals, $I$, the following procedure is used: 
\begin{enumerate}
\item Express $I_{a_1 b_1; \cdots ; a_m b_m}$ as a linear combination of appropriate delta function products, with unknown coefficients, $x_i$. \label{step1}
\item Simplify this expression, taking into account symmetries of the integral with respect to $\{a_1,\ldots ,a_m\}$ and $\{b_1,\ldots ,b_m\}$. Some $x_i$ will be found to be equal to others, reducing the number of independent $x_i$. For example, from \eqn{i3defn}, we see that $I_{ab;a'b';a''b''} = I_{a''b'';a'b';ab}$. Making use of this and \eqn{i3lincomb} we find that $x_2 = x_3$ and $x_4 = x_5$. 
\item Construct a system of equations for $x_i$ in terms of known integrals by contracting appropriate combinations of indices.
\item Solve this system for $x_i$ and substitute into the original expression for $I_{a_1 b_1; \cdots ; a_m b_m}$ from step (\ref{step1}).
\end{enumerate}

\section{Integrals Required for Moment Calculations}
\label{momentints}
In this section we calculate a number of integrals useful in Hamiltonian LGT, especially for those techniques requiring the calculation of Hamiltonian moments. The calculations make use of the results of \sect{creutzextension}.\\

The Hamiltonian methods for LGT generally require the calculation of various integrals over SU($N$). The $t$-expansion~\cite{Horn:1985ax} and plaquette expansion~\cite{Hollenberg:1993bp} methods are two approaches to Hamiltonian LGT which require the calculation of Hamiltonian moments. These moments can be simplified so that all electric field operators are removed, and thus reduced to a set of group integrals over SU($N$). For simple one-plaquette trial states in 2+1 dimensions these integrals can be calculated analytically. More complicated scenarios (3+1 dimensions or more complicated trial stated) typically require numerical evaluation.\\

In the calculation of Hamiltonian moments in 2+1 dimensions, the following integrals require calculation at the fourth order for diagrams on two-plaquette skeletons (in the following $U$ and $V$ represent the neighbouring plaquettes of the two-plaquette connected skeleton):
\bea
Z_1(c,N) &=& \la \phi_0' |\trr (UV) \trr (UV)|\phi_0' \ra \equiv 
\left\la \rectangleAA \right\ra\la \phi_0|\phi_0 \ra \label{Z1} \\
Z_2(c,N) &=& \la \phi_0' |\trr (UV) \trr (UV^\dagger)|\phi_0'\ra \equiv  \left\la \rectangleAC \right\ra \la \phi_0|\phi_0 \ra \label{Z2} \\
Z_3(c,N) &=& \la \phi_0' | \trr (UV^\dagger) \trr (UV^\dagger)| \phi_0' \ra \equiv  \left\la \rectangleAD \right\ra\la \phi_0|\phi_0 \ra \label{Z3} \\
Z_{4}(c,N) &=& \la  \phi_0' |\trr (UV) \trr (U^\dagger V^\dagger)| \phi_0'\ra \equiv  \left\la \rectangleAAd \right\ra\la \phi_0|\phi_0 \ra \label{Z4}\\
Z_5(c,N) &=& \la  \phi_0' |\trr (UUVV)| \phi_0'\ra  \equiv  \left\la \rectangleAE \right\ra \la \phi_0|\phi_0 \ra\\
Z_6(c,N) &=& \la  \phi_0' |\trr (UUV^\dagger V^\dagger)| \phi_0'\ra \equiv  \left\la \rectangleAF \right\ra\la \phi_0|\phi_0 \ra \\
Z_7(c,N) &=& \la  \phi_0' |\trr (UVUV)| \phi_0'\ra \equiv  \left\la \rectangleAB \right\ra \la \phi_0|\phi_0 \ra \label{Z7} \\
Z_8(c,N) &=& \la \phi_0' | \trr (UV^\dagger UV)| \phi_0'\ra  \equiv \left\la \rectangleAG \right\ra\la \phi_0|\phi_0 \ra \\
Z_9(c,N) &=& \la \phi_0' | \trr (UV^\dagger U^\dagger V)| \phi_0'\ra \equiv  \left\la \rectangleAH \right \ra\la \phi_0|\phi_0 \ra \\
Z_{10}(c,N) &=& \la \phi_0' | \trr (UV^\dagger UV^\dagger)| \phi_0'\ra \equiv  \left\la \rectangleAI \right \ra\la \phi_0|\phi_0 \ra \label{Z10}
\eea
Here each integral is calculated over SU($N$) and $|\phi_0\ra$ is the one plaquette trial state:
\be
 |\phi_0 \ra = \exp \left[ \frac{c}{2}\sum_p \trr(U_p+U_p^\dagger)\right]|0\ra,
\ee
where $|0\ra$ is the strong coupling vacuum and the sum is over all plaquettes on the lattice. $|\phi_0'\ra$ is the reduced trial state incorporating only the neighbouring plaquettes $U$ and $V$,
\be
|\phi_0' \ra = \exp \left[ \frac{c}{2}\trr(U+U^\dagger+V+V^\dagger)\right]|0\ra.
\ee

Eqs.~(\ref{Z1}), (\ref{Z3}), (\ref{Z4}), (\ref{Z7}) and (\ref{Z10}) can be calcu\label{J3}lated using standard character expansion techniques~\cite{Carlsson:2003wx}. Here we will calculate these integrals using a different method which is able to be applied to the calculation of the other integrals presented above. We start with $Z_1(c,N)$, which can expressed as (assuming summation over repeated indices):
\bea
Z_1(c,N) &= \int dU dV e^{c\trr(U+U^\dagger)}e^{c\trr(V+V^\dagger)}U_{ab}V_{ba}U_{a'b'}V_{b'a'}\nn\\
 &= I_{ab;a'b'}(c) I_{ba;b'a'}(c).\label{Z1calc}
\eea
Making use of \eqns{lincomb}{b2}, $Z_1$ can be reduced to an expression involving known single plaquette integrals as follows:
\bea
Z_1(c,N)&=& (b_1 \delta_{ab}\delta_{a'b'} + b_2 \delta_{ab'}\delta_{ba'})(b_1 \delta_{ba}\delta_{b'a'} + b_2 \delta_{ba'}\delta_{ab'}) \\
&=& N^2 b_1^2 + 2 N b_1 b_2 + N^2 b_2^2 \\
&=& \frac{N I_{aa;bb}^2- 2 I_{aa;bb}I_{ab;ba}+N I_{ab;ba}^2}{N(N^2-1)}
\eea

Next we calculate $Z_2$. Using the same technique we have,
\bea
Z_2(c,N) &=& \int dU dV e^{c\trr(U+U^\dagger)}e^{c\trr(V+V^\dagger)}U_{ab}V_{ba}U_{a'b'}(V^\dagger)_{b'a'} \\
&=& I_{ab;a'b'}(c) J_{ba;b'a'}(c).\label{Z2calc}
\eea
Making use of \eqns{lincomb}{b2} and also \eqns{Jindels}{Jdetail}, $Z_2$ can be reduced to an expression involving known integrals as follows,
\bea
Z_2(c,N) &=&  I_{ab;a'b'}(c) J_{ba;b'a'}(c)\\
&=& N^2 b_1 d_1 + N b_1 d_2 + N b_2 d_1 + N^2 b_2 d_2 \\
&=& \frac{N I_{aa;bb}J_{aa;bb} - I_{aa;bb}J_{ab;ba}- I_{ab;ba}J_{aa;bb}+ N I_{ab;ba}J_{ab;ba}}{N(N^2-1)}.
\eea
$Z_3$ can be shown to be equal to $Z_1$ as follows,
 \bea
Z_3(c,N) &= \int dU dV e^{c\trr(U+U^\dagger)}e^{c\trr (V+V^\dagger)}U_{ab}U_{cd}(V^\dagger)_{ba}(V^\dagger)_{dc} \nn\\
&= I_{ab;cd}(c) I_{ba;dc}(c) = Z_1(c,N).\label{Z3calc} 
\eea
Here the last line follows from the fact that an integral of a single plaquette loop does not depend on its direction. 

The remainder of the integrals can be calculated similarly. The results are as follows:
\bea
Z_{4}(c,N) &=& \frac{1}{N^2-1} ( J_{aa;bb}^2+  J_{ab;ba}^2) - \frac{2}{N(N^2-1)}J_{ab;ba}J_{aa;bb}\\
Z_5(c,N) &=&  \frac{1}{N} I_{ab;ba}^2 \\
Z_6(c,N) &=& Z_5(c,N) \\
Z_7(c,N) &=& \frac{2 N I_{aa;bb}I_{ab;ba}-I_{aa;bb}^2 -I_{ab;ba}^2}{N(N^2-1)} \\
Z_8(c,N) &=& \frac{N I_{aa;bb}J_{ab;ba}-I_{aa;bb}J_{aa;bb} - I_{ab;ba}J_{ab;ba}+N I_{ab;ba}J_{aa;bb}}{N(N^2-1)} \\
Z_9(c,N) &=& \frac{2 N J_{aa;bb}J_{ab;ba}-J_{aa;bb}^2 - J_{ab;ba}^2}{N(N^2-1)} \\
Z_{10}(c,N) &=& Z_6(c,N)
\eea
The techniques presented in this section can be extended to the calculation of loop integrals on three or more neighbouring plaquettes using \eqns{I3}{JJ3} and their extensions.

\section{Conclusion}
In this paper a number of integrals over SU($N$) have been calculated. Further work could incorporate these calculations into plaquette expansion calculations of glueball masses in 2+1 dimensions. Other calculations involving Hamiltonian moments may be able to be extended to more complicated trial states, at least in 2+1 dimensions, using the techniques presented in this paper.

\section*{References}
\bibliographystyle{unsrt} 
\bibliography{intpaper}

\end{document}